\documentclass[aps,pra,showkeys,twocolumn,superscriptaddress]{revtex4-2}
\usepackage{array}
\usepackage{booktabs}
\usepackage{tabu}
\usepackage{dcolumn}
\usepackage{amsmath}
\usepackage{amsfonts}
\usepackage{float}
\usepackage{amssymb}
\usepackage{graphicx,color}
\usepackage[colorlinks={true}]{hyperref}
\hypersetup{colorlinks=true,linkcolor=red,citecolor=blue,urlcolor=blue}
\usepackage{graphicx}
\usepackage{subfigure}
\usepackage{graphicx}
\usepackage{dcolumn}
\usepackage{bm}
\usepackage{pstricks}
\usepackage{braket}

\def\be{\begin{equation}}
  \def\ee{\end{equation}}
\def\bea{\begin{eqnarray}}
\def\eea{\end{eqnarray}}
\def\f{\frac}
\def\n{\nonumber}
\def\l{\label}
\def\p{\phi}
\def\o{\over}
\def\R{\rho}
\def\pa{\partial}
\def\om{\omega}
\def\na{\nabla}
\def\P{\Phi}
\begin{document}
\title{Thermal local quantum uncertainty in a two-qubit-superconducting system under decoherence}
\author{M. R. Pourkarimi}\email{Corresponding author: mrpourkarimy@gmail.com}
\affiliation{Department of Physics, Salman Farsi University of Kazerun, Kazerun, Iran}
\author{S. Haddadi}\email{Corresponding author: saeed@ssqig.com}
\affiliation{Faculty of Physics, Semnan University, P.O.Box 35195-363, Semnan, Iran}
\author{M. Nashaat}
\affiliation{Department of Physics, Faculty of Science, Cairo University, 12613, Giza, Egypt}
\affiliation{BLTP, JINR, Dubna, Moscow Region, 141980, Russia}
\author{K. V. Kulikov}
\affiliation{BLTP, JINR, Dubna, Moscow Region, 141980, Russia}
\affiliation{Dubna State University, Dubna, Moscow Region, 141980, Russia}
\author{Yu. M. Shukrinov}
\affiliation{BLTP, JINR, Dubna, Moscow Region, 141980, Russia}
\affiliation{Dubna State University, Dubna, Moscow Region, 141980, Russia}
\affiliation{Moscow Institute of Physics and Technology, Dolgoprudny, 141700, Russia}

\date{\today}
\def\be{\begin{equation}}
  \def\ee{\end{equation}}
\def\bea{\begin{eqnarray}}
\def\eea{\end{eqnarray}}
\def\f{\frac}
\def\n{\nonumber}
\def\l{\label}
\def\p{\phi}
\def\o{\over}
\def\R{\rho}
\def\pa{\partial}
\def\om{\omega}
\def\na{\nabla}
\def\P{$\Phi$}

\begin{abstract}
By considering the local quantum uncertainty (LQU) as a measure of quantum correlations, the thermal evolution of a two-qubit-superconducting system is investigated.  We show that the thermal LQU can be increased by manipulating the Hamiltonian parameters such as the mutual coupling and Josephson energies, however, it undergoes sudden transitions at specific temperatures. Furthermore, a detailed analysis is presented regarding the impact of decohering channels on thermal LQU. This controllable LQU in engineering applications can disclose the advantage enabled in the superconducting charge qubits for designing quantum computers and quantum batteries.
\end{abstract}


\maketitle

\section{INTRODUCTION}	
Quantum correlations such as quantum entanglement are a central part in quantum information science \cite{Fedorova2022}. It has been manifested in the Einstein–Podolsky–Rosen thought,
Bell’s inequality test, quantum cryptography, etc. \cite{Gisin2002,Amico2008,Horodecki2009,Reid2009}. Depending on whether the nature of the environments is Markovian or non-Markovian \cite{Artur1},  dynamics of quantum entanglement may present some phenomena such as sudden death, birth or trapping \cite{Guo2019,Francesco2011,Zhou2009,Bellomo2008,Dajka2008,Yu2004,Artur2}.

The authors of Refs. \cite{ Ferraro2010,Modi2012} have shown that quantum correlation and quantum entanglement should be treated separately, and demonstrated that entanglement is not the only type of correlation for the implementation of quantum protocols, with some separable states may also perform better than their classical counterparts. This leads to new quantifiers which are capable of detecting non-classical correlations beyond entanglement \cite{Modi2012, Dakic2010}, and develop new proposals for quantum correlation measurements \cite{Adesso2016,Bera2018,dwang}.

An important question arises as to whether individual measures ought to be ascribed to their respective correlations, or alternatively, it should be contended that a singular quantum correlation exists, yet the techniques employed to describe it are not sufficiently flawless. Today, the understanding of this question is that the quantum correlation is one, however, various methods have been proposed to quantify this correlation, each of which possesses notable imperfections \cite{Fedorova2022}.

From this point of view, the authors of Refs. \cite{Zurek2000,Ollivier2001} formulated the notion of quantum discord, which quantifies the presence of quantum correlations between two subsystems within a quantum system. Incorporating further to the discourse, the authors in Refs. \cite{Henderson2001,Vedral2003} posited a measure for the exclusively classical correlation, which, upon its deduction from the entire correlation, arrived at an equivalent quantum correlation as the discord \cite{Fedorova2022}. The discord scales with the quantum efficiency, whereas the degree of entanglement remains negligibly small during the entire computation. These intriguing discoveries warrant scrutiny towards the novel measure of quantum correlation \cite{Merali2011,Datta2008,Modi2012,Aldoshin2014,Streltsov}.

Several theoretical and experimental studies of the different quantum systems have demonstrated that quantum entanglement and quantum discord may have differences in quantitative and even qualitative behaviors \cite{Werlang2010,Guo2011,Campbell2013}. For pure quantum states, the quantum discord and entanglement are the same. However, for mixed states, quantum discord can exist when quantum entanglement is identically equal to zero \cite{Zyczkowski98,Ferraro2010}.  In other words, quantum discord and entanglement exhibit distinct characteristics when evaluating the simplest mixed states such as Bell-diagonal and Werner states \cite{Moreva2017}.

The local quantum uncertainty (LQU) \cite{Girolami} is an additional form of quantum correlations which solely captures the quantum correlations, without their classical counterpart. LQU, as a discord-like measure for
evaluating quantum correlations, can be defined as the minimum value of the skew information \cite{Tian2020}. In comparison with quantum discord, LQU  does not need a complicated optimization procedure over the measurements. Several works considered LQU, e.g., in quantum phase transitions \cite{Karpat2014,Coulamy2016}, its relationship with quantum Fisher information \cite{Wu2018,hadpre,dahbi,bena,Yurischev1,Yurischev2}, for three qubits systems \cite{Slaoui1,Slaoui2,Slaoui3,Slaoui4}, and so on \cite{Jebli2017,Habiballah2018,Chen2020,hadsci,obada,rahman2022,Khedr}.

Structures based on the different types of Josephson junctions present an interesting object for fundamental research and wide applications in quantum metrology, superconducting electronics and spintronics \cite{Golubov2017,linder2015}. Intrinsic Josephson effect in high-temperature superconductors demonstrates the coupling between JJs formed by superconducting layers, multiple branch structures in IV-characteristics and reach resonance features and serve as a source of coherent terahertz radiation \cite{Shukrinov-sust2017,welp2013,shukrinov-prl2007}. Novel features and applications were found recently in SFS JJ such as anomalous Josephson effect \cite{shukrinov-ufn2022,mazanik-pra2020,shukrinov-prb2021,abdelmoneim-prb2022,linder2015}, superconducting diode effect \cite{linder-condmat}, and phase batteries \cite{strambini2020}.
Superconducting quantum circuits, which are composed of Josephson junctions, are considered to be macroscopic circuits that have the potential to exhibit quantum mechanical behavior similar to that of artificial atoms. This unique characteristic enables researchers to investigate quantum entanglement and quantum coherence on a macroscopic scale, as evidenced by several studies \cite{You2005,Schoelkopf2008,You2011,Xiang2013,Georgescu2014,Dong2015}. Additionally, these synthetic atoms offer a promising avenue for implementing quantum information technology and for testing the laws of quantum mechanics on larger systems \cite{Dong2015}.
Entanglement between different kinds of Josephson junction qubits has been studied theoretically and experimentally \cite{Tang2022,Tian2010,McDermott2005,Wallraff2004,Pashkin2003}, e.g., in charge \cite{Bahrova2021,Contreras2008,Yamamoto2003}, flux \cite{Yamamoto2005,Vion2002,alx2023} and phase \cite{Martinis2002,Berkley2003,McDermott2005} qubits.

The current model is constituted of two singular Cooper-pair box charge qubits that are interlinked via a fixed capacitor. The coupling energy that exists between the aforementioned qubits can be effectively utilized to estimate the degree of entanglement for the two-qubit system. The Cooper-pair box may host a two-level system, namely $\ket{0}$ and $\ket{1}$. These states differ by $2e$ ($e$ is the electronic charge) from a Cooper pair and are coherently superimposed by the Josephson coupling \cite{Yamamoto2003,Bouchiat1998}. Connecting the two Cooper-pair boxes through a capacitor yields interference of the quantum states, which ultimately leads to quantum beating, as posited by Yamamoto in  \cite{Yamamoto2003}. The authors of Ref. \cite{Yamamoto2003} leveraged the coherent system comprised of the charge states $\ket{00}$, $\ket{01}$, $\ket{10}$, and $\ket{11}$, to showcase the implementation of a logic gate. Furthermore, the authors demonstrated that this logic gate functions seamlessly as a quantum gate.

Motivated by the above issues, we use LQU for the theoretical investigation of quantum correlations measurement in a two-qubit-superconducting (TQS) system that is at a thermal regime.  We study the temperature dependence of LQU at different values of Josephson and coupling energies. A maximum on this dependence is observed, which is explained through the temperature dependence of the largest eigenvalue of a symmetric matrix determining LQU. Besides, the effect of Hamiltonian parameters and decohering channels on thermal LQU is discussed in detail.
Hence, this paper is categorized as follows. In Sec. \ref{sec:2}, we present LQU as a measure of quantum correlations and we disclose the physical model considered for the TQS system in Sec. \ref{sec:3}.  Next, we provide the main results in Sec. \ref{sec:4} and finally, we summarize them in Sec. \ref{sec:5}.

\section{Local quantum uncertainty}\label{sec:2}

In the literature, many quantum criteria have been introduced for measuring quantum correlations such as concurrence, quantum discord and so on. As mentioned before, LQU is a measure of quantum correlations based on quantum uncertainty \cite{Girolami}.
In quantum world, two incompatible observables cannot be measured with arbitrary accuracy as presented in Heisenberg uncertainty relation. However, because of intrinsic uncertainty in quantum measurement, quantum uncertainty can be shown for the measurement of a single observable.  For a bipartite quantum correlated state, there is a quantum uncertainty when measuring a subsystem locally.  This happens because, a quantum correlated state cannot be the eigenstate of the local operators.

For a bipartite quantum system with density matrix $\rho_{AB}$, LQU can be obtained from skew information by minimizing over all local Hermitian operator $K_{A}$ as follows \cite{khed2019,khed2021}
\begin{equation}
\label{lqu}
\textmd{LQU}(\rho_{AB}):=\min_{K_A}\mathcal{I}(\rho_{AB},K_A\otimes I_B),
\end{equation}
where $K_{A}$ being the local observable with a non-degenerate spectrum, $\mathcal{I}\left(\rho_{AB}, K_A \otimes I_B\right)=-\frac{1}{2} \operatorname{Tr}\left\{\left[\sqrt{\rho_{AB}}, K_A \otimes I_B\right]^2\right\}$
is the skew information, $[., .]$ denotes the commutator, and $I_B$ is the identity operator acting on the subsystem $B$.

For a bipartite $2\otimes d $ state when the subsystem $A$ is a single qubit, LQU can be defined as the simple following form \cite{zidan2023}
\begin{equation}\label{lqu2}
\textmd{LQU}(\rho_{AB})=1- \lambda_{max}(W_{AB}),
\end{equation}
where $\lambda_{max}(W_{AB})$ is the largest eigenvalue of a symmetric $3\times 3$ matrix $W_{AB}$, whose elements read \cite{Girolami}
\begin{equation}
\label{wab}
(W_{AB})_{ij}=\operatorname{Tr} \left \{\sqrt{ \rho_{AB} }(\sigma_{A i}\otimes I_B)\sqrt{ \rho_{AB} }(\sigma_{A j}\otimes I_B)  \right\},
\end{equation}
where $\sigma_{A i,j}$ $(i,j\in\{x,y,z\})$ are Pauli operators acting on subsystem $A$.

\section{Theoretical model and thermalization}\label{sec:3}
The Hamiltonian of a TQS system is given by \cite{Shaw2009,Tian2011,zidan1,zidan2,rinp2023,mansourm2023}
\begin{align} \label{MH2}
\mathcal{H}_{TQS}:=-\frac{1}{2}\left[E_{j_1}\sigma_{x_1}+ E_{j_2}\sigma_{x_2}-2E_{m}\sigma_{zz}\right],
\end{align}
where $E_{j_1}$ and $E_{j_2}$  denote Josephson energies and $E_m$ is the mutual coupling energy. Moreover, $\sigma_{x_1}=\sigma_{x}\otimes I$,  $\sigma_{x_2}=I \otimes \sigma_{x}$ and $\sigma_{zz}=\sigma_{z}\otimes \sigma_{z}$ in which $\sigma_{x,z}$ are the Pauli operators. This Hamiltonian by assuming the equal Josephson energy splitting for both qubits $E_{j_{1}}=E_{j_{2}}$ $=E_{j}$ can be written as follows
\begin{equation}
\mathcal{H}_{TQS}=-\frac{1}{2}\left(
\begin{array}{cccc}
-2E_{m} & E_{j} & E_{j} & {0} \\
E_{j} & 2E_{m} & {0} & E_{j} \\
E_{j} & {0} & 2E_{m} & E_{j} \\
{0} & E_{j} & E_{j} & -2E_{m}%
\end{array}%
\right).
\end{equation}

Without loss of generality, by using some local unitary orthogonal transformations (the Hadamard--$\textmd{H}$ transform of the Pauli matrices, i.e. $\textmd{H}\sigma_{x}\textmd{H}=\sigma_{z}$ and $\textmd{H}\sigma_{z}\textmd{H}=\sigma_{x}$ where $\textmd{H}:=(\sigma_{x}+\sigma_{z})/\sqrt{2}$ is Hadamard
gate), our new Hamiltonian can be expressed as
\begin{equation}\label{MH}
\mathcal{H}\equiv \mathcal{H}_{new}=\left(
\begin{array}{cccc}
-E_{j} & 0 & 0 & E_{m} \\
0 & 0 & E_{m} & 0 \\
0 & E_{m} & 0 & 0 \\
E_{m} & 0 & 0 & E_{j}%
\end{array}%
\right).
\end{equation}

When a typical system with Hamiltonian $\mathcal{H}$ is in the thermal equilibrium at temperature $T$, its density operator may be described as
\begin{equation}
\label{Eq13}
\varrho_T=\frac{1}{Z}\exp(-\beta \mathcal{H}),
\end{equation}
where $Z=\operatorname{Tr}[\exp( -\beta \mathcal{H})]$ is the partition function of the system, and $\beta=1/k_{B}T$ in which $k_{B}$ is the Boltzmann constant, however, we set $k_{B}=1$ for simplicity. Hence, the thermal density matrix (\ref{Eq13}) based on our Hamiltonian (\ref{MH}) has the following X form
\begin{equation}\label{thermal}
\varrho_T=\left(
\begin{array}{cccc}
 a^{+} & 0 & 0 & c \\
 0 & b & d & 0 \\
 0 & d & b & 0 \\
 c & 0 & 0 & a^{-} \\
\end{array}
\right),
\end{equation}
where
\begin{align}
&a^{\pm}=\frac{1}{Z}\left[\cosh \left(\alpha/T\right)\pm\frac{E_j \sinh \left(\alpha/T\right)}{\alpha}\right],\nonumber\\
&b=\frac{1}{Z}\left[\cosh \left(E_m/T\right)\right],\nonumber\\
&c=\frac{1}{Z}\left[-\frac{E_m \sinh \left(\alpha/T\right)}{\alpha}\right],\nonumber\\
&d=\frac{1}{Z}\left[-\sinh \left(E_m/T\right)\right],
\end{align}
with
$Z=2 \left[\cosh \left(\alpha/T\right)+\cosh \left(E_m/T\right)\right]$ and
$\alpha=\sqrt{E_{m}^2+E_{j}^2}$.
Using now the general formulas from \cite{Yurischev2}, one can obtain the eigenvalues of the matrix \eqref{wab} according to thermal state \eqref{thermal} as
\begin{align}
W_{11}:=\lambda_1=&\left(\sqrt{\gamma_1}+\sqrt{\gamma_2}\right)\left(\sqrt{\gamma_3}+\sqrt{\gamma_4}\right)\nonumber\\
&+\frac{4|c d|}{\left(\sqrt{\gamma_1}+\sqrt{\gamma_2}\right)\left(\sqrt{\gamma_3}+\sqrt{\gamma_4}\right)}, \nonumber\\
W_{22}:=\lambda_2=&\left(\sqrt{\gamma_1}+\sqrt{\gamma_2}\right)\left(\sqrt{\gamma_3}+\sqrt{\gamma_4}\right)\nonumber\\
&-\frac{4|c d|}{\left(\sqrt{\gamma_1}+\sqrt{\gamma_2}\right)\left(\sqrt{\gamma_3}+\sqrt{\gamma_4}\right)},\nonumber\\
W_{33}:=\lambda_3=&\frac{1}{2}\bigg[\left(\sqrt{\gamma_1}+\sqrt{\gamma_2}\right)^2+\left(\sqrt{\gamma_3}+\sqrt{\gamma_4}\right)^2
\nonumber\\
&+\frac{(a^{-}-a^{+})^2 - 4|c|^2}{\left(\sqrt{\gamma_1}+\sqrt{\gamma_2}\right)^2}-\frac{4|d|^2}{\left(\sqrt{\gamma_3}+\sqrt{\gamma_4}\right)^2}\bigg],
\end{align}
with $\gamma_{1,2}=(a^{+}+a^{-} \pm \sqrt{(a^{+}-a^{-})^2 + 4|c|^2})/2$ and $\gamma_{3,4}=b \pm |d|$. Knowing that $\lambda_1 \geq \lambda_2$, an explicit
expression for LQU \eqref{lqu2} is expressed as

\begin{equation}\label{lqu4}
\textmd{LQU}(\varrho_T)=1- \max\{\lambda_1,  \lambda_3\}.
\end{equation}

\begin{figure}[t]
   \centering
  \includegraphics[width=0.45\textwidth]{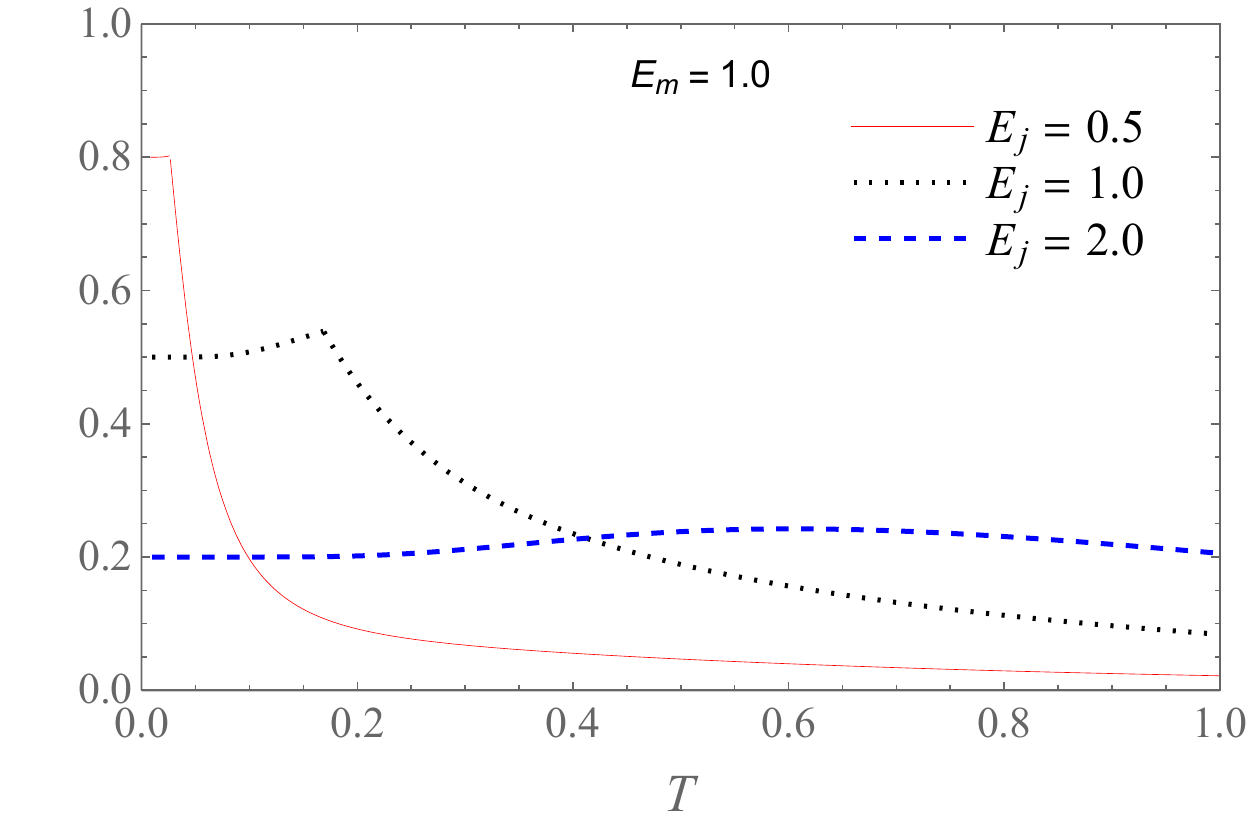}(a)\\
  \includegraphics[width=0.45\textwidth]{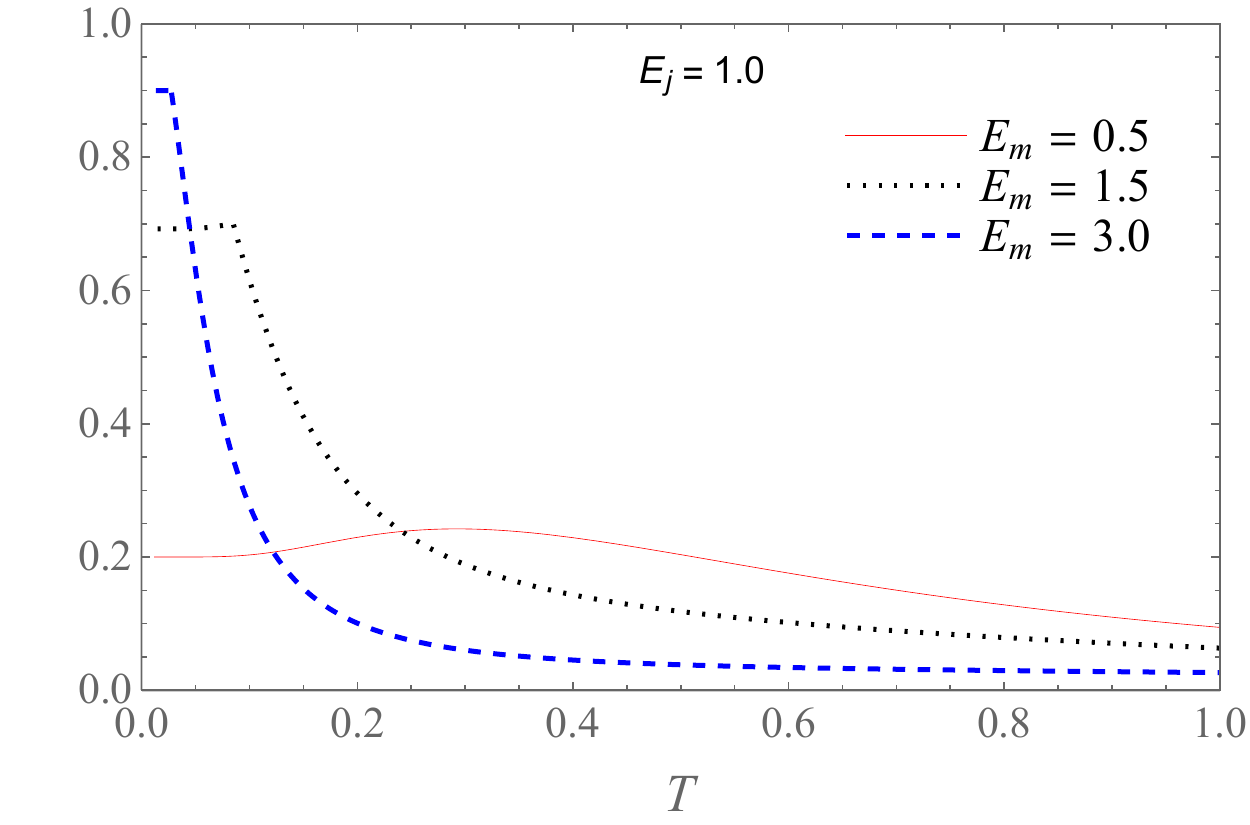}(b)
\caption{LQU versus temperature $T$ for the three fixed values of (a) $E_j$ and (b) $E_m$.}
\label{fig:1}
\end{figure}

\begin{figure}[htb]
   \centering
  \includegraphics[width=0.45\textwidth]{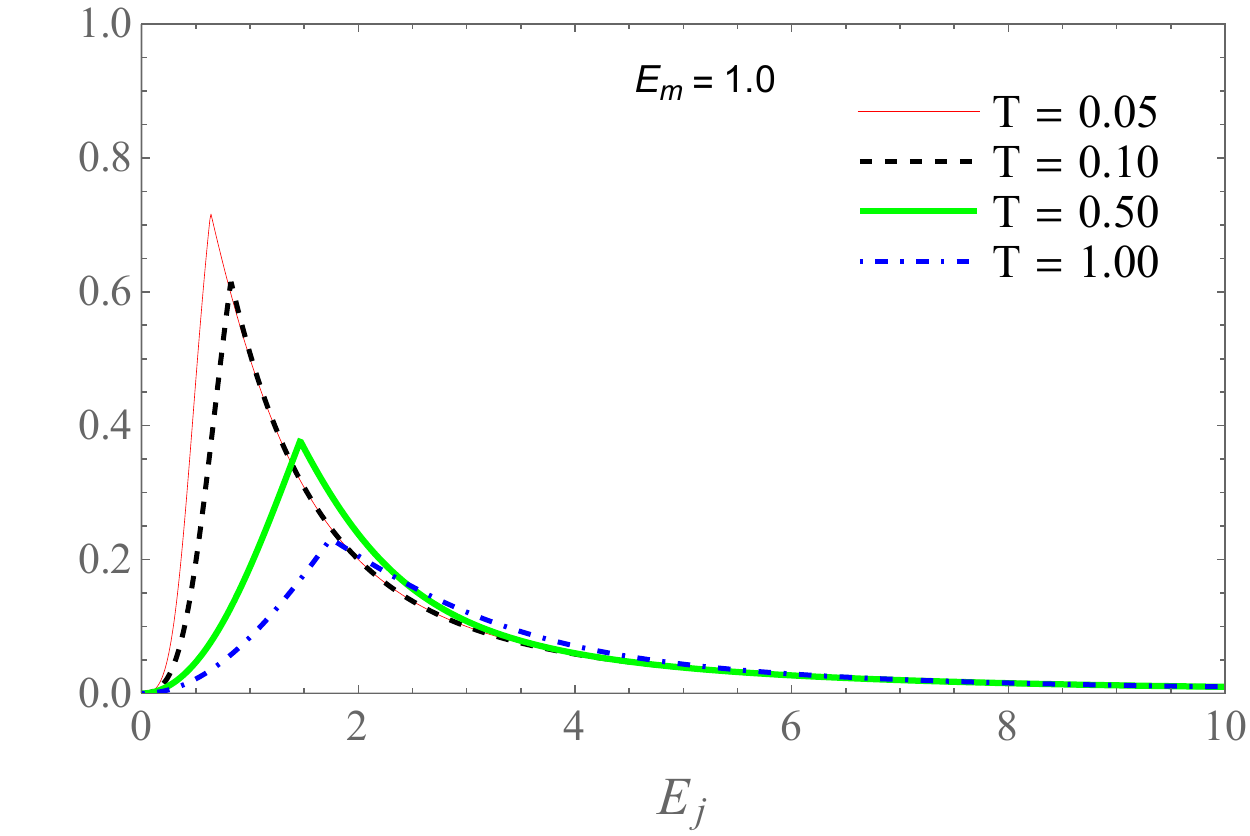}(a)\\
  \includegraphics[width=0.45\textwidth]{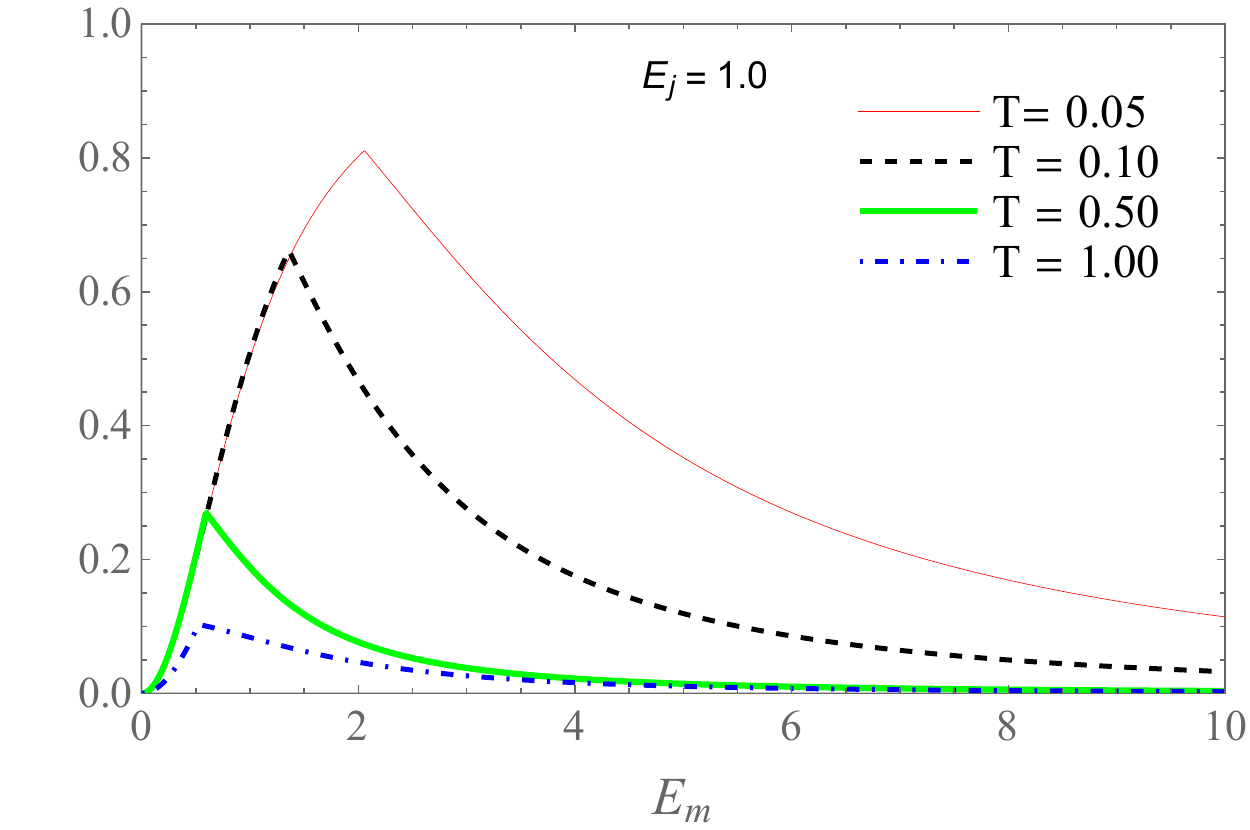}(b)
\caption{LQU versus (a) $E_j$ and (b) $E_m$ for the four fixed values of temperature $T$.}
\label{fig:2}
\end{figure}

\section{Main results and discussion}\label{sec:4}

\subsection{Thermal LQU in a TQS system}
One of the important points in fundamental research and the application of quantum correlations is finding the possibility to manipulate and control different types of quantum criteria and measures. Here, we present the simulation results of  LQU and demonstrate the effects of temperature $T$, Josephson energy $E_j$, and coupling energy $E_m$ on it. As we expect, the LQU peaks fall to zero by increasing $T$. Its physical justification is that the ground state is mixed with the excited states at high
temperatures.

The LQU temperature dependence at different Josephson energies and fixed coupling energy is shown in figure \ref{fig:1}(a), while in figure \ref{fig:1}(b) we show it for fixed Josephson energy.
By increasing the temperature, the LQU shows a sharp decrease after $T\approx0.05$ for $E_{j}=0.5$ at $E_{m}=1$ as shown in figure \ref{fig:1}(a), while for $E_{j}=2$ it changes smoothly around 0.2 with valley shape before reaching a stable value at high temperature. Notice, the LQU undergoes a sudden transition after $T\approx0.15$ for $E_{j}=1$.
Next, we demonstrate the LQU temperature dependence at fixed Josephson energy $E_j=1$ in figure \ref{fig:1}(b). At low temperatures, LQU can be increased by increasing coupling energies, but at high temperatures, this scenario is completely opposite.

The LQU energy dependence at different temperatures is illustrated in figures \ref{fig:2}(a) and \ref{fig:2}(b).
As expected from Eq. \eqref{thermal}, the off-diagonal elements of the thermal density matrix $\varrho_T$ are zero at $E_{j}=E_{m}=0$. Therefore, no quantum correlation exists since we have $\varrho_T=\textmd{diag}\{1/4,1/4,1/4,1/4\}$, which is in accordance with the outcome obtained in Ref. \cite{Shaw2009} for the quantum entanglement captured by the concurrence.
However, by increasing $E_{j}$ or $E_{m}$, the LQU increases sharply and reaches a maximum value, then it decays to a constant value.  Interestingly, the qualitative behavior of thermal LQU is in good agreement with the experimental data for thermal entanglement as reported in Refs. \cite{Shaw2009,Tian2011}. Notice, the maximum of the LQU occurs at a low temperature and lower Josephson energy [see figure \ref{fig:2}(a)]. While it occurs at higher coupling energy, as seen in figure \ref{fig:2}(b).

\begin{figure}[htb]
	\centering
	\includegraphics[width=1\linewidth]{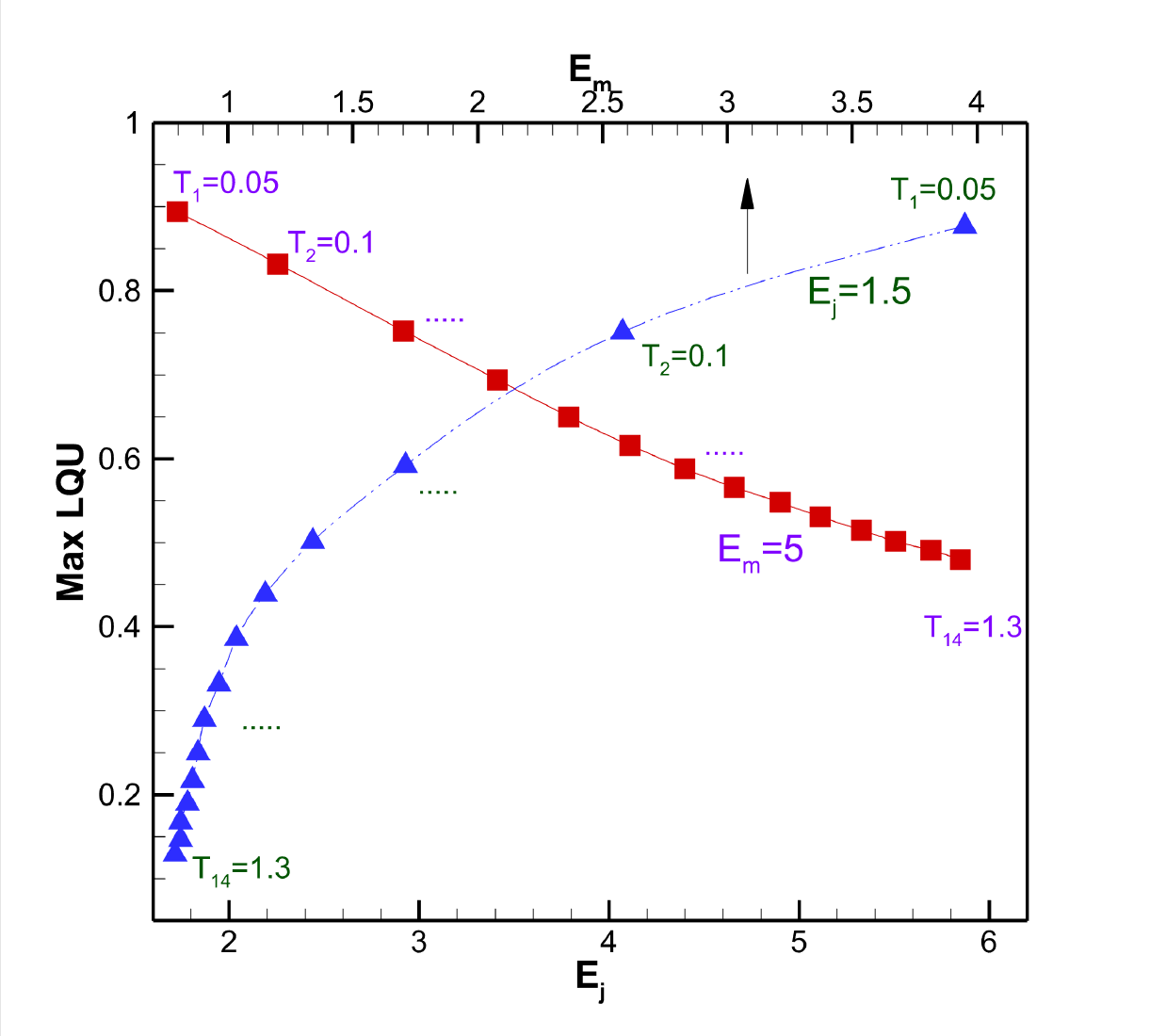}
	\caption{Maximum value of the LQU at different temperatures. The red curve is recorded for $E_{m}=5$, while the blue one is recorded for $E_{j}=1.5$. }
	\label{fig:3}
\end{figure}

To compare in more detail, the thermal evolution of LQU is shown in figure \ref{fig:3}. Herein, we plot the maximum value of LQU at different temperatures versus the Josephson and coupling energies. One can clearly find the difference between the effects of Josephson energy and coupling energy. As can be seen from this figure, the maximum LQU at $T=0.05$ is achieved for $E_{j}=1.728$, while it occurs for $E_{m}=3.95$, i.e., Josephson energy and coupling energy dependence of the ``Max LQU" is opposite to each other.
Generally, by increasing the temperature, the ground state and excited states are mixed, and the LQU decreases. This decrease is edgy at larger coupling energy and smaller Josephson energy.  For the higher Josephson energy, the changes in the LQU become smooth till it reaches a constant value. This means that the LQU is sensitive to Josephson energy at lower temperatures and higher coupling energies.

\begin{figure}[htb]
	\centering
		\includegraphics[width=0.95\linewidth]{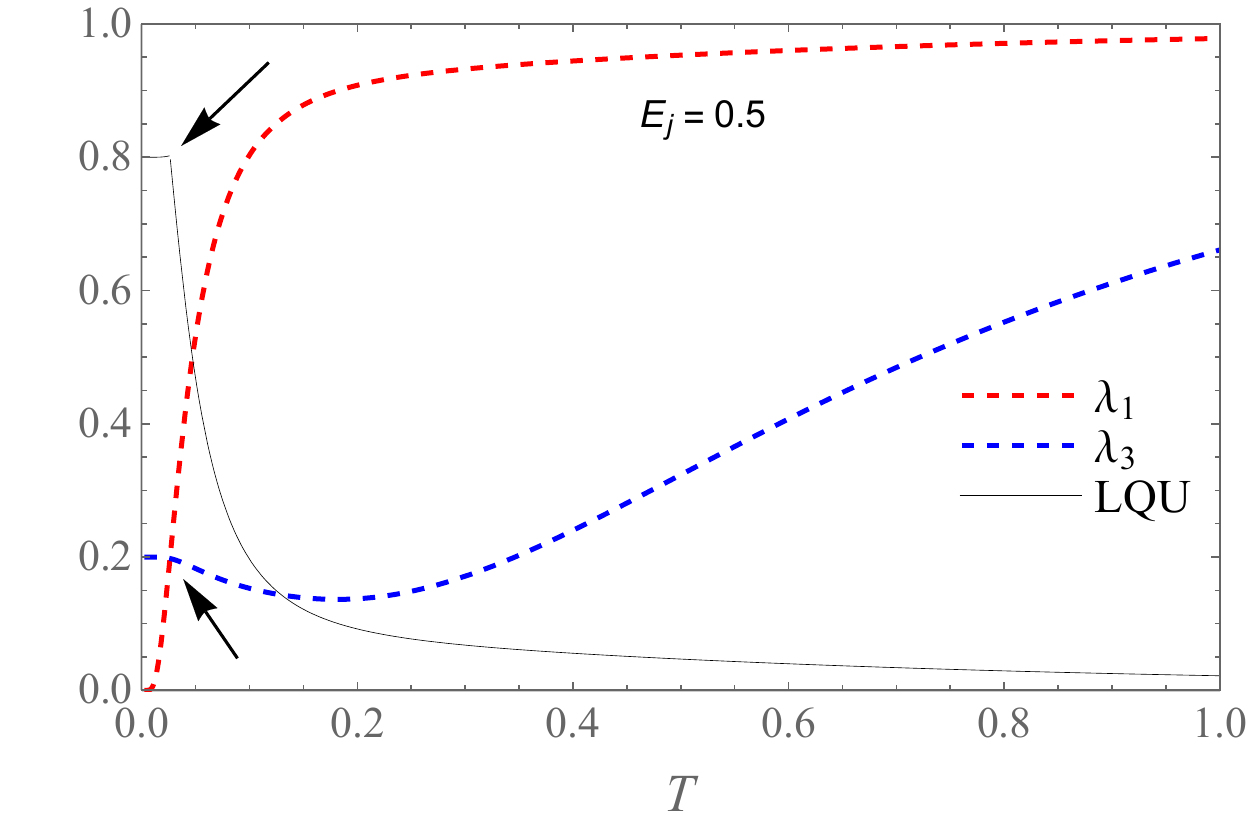}(a)\\
		\includegraphics[width=0.95\linewidth]{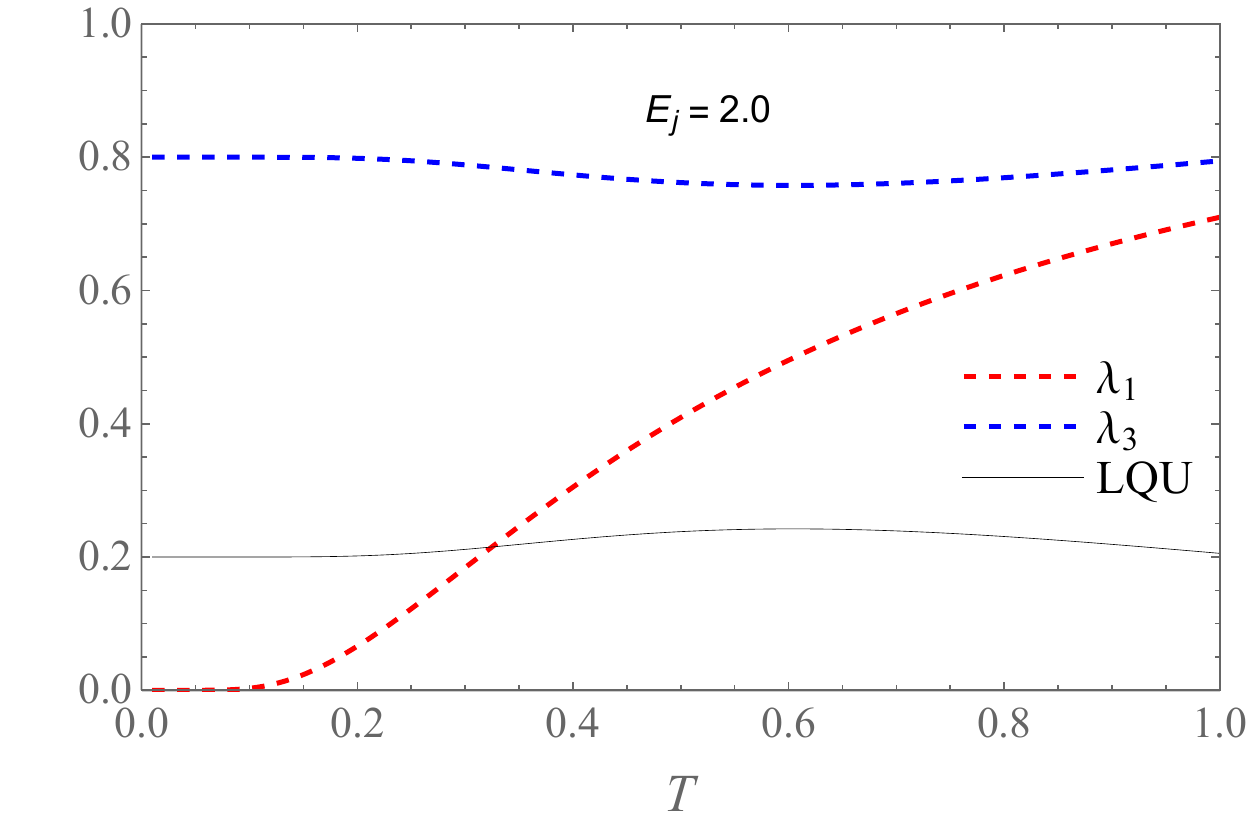}(b)
	\caption{LQU and temperature dependence of the eigenvalues $\lambda_{1}$ and $\lambda_{3}$ for (a) $E_{j}=0.5$ and (b) $E_{j}=2$ with $E_{m}=1$. The downward arrow indicates the sudden change point of LQU and the upward arrow shows the intersection point between $\lambda_{1}$ and $\lambda_{3}$.}
	\label{ej01}
\end{figure}

\begin{figure}[htb]
	\centering
		\includegraphics[width=0.95\linewidth]{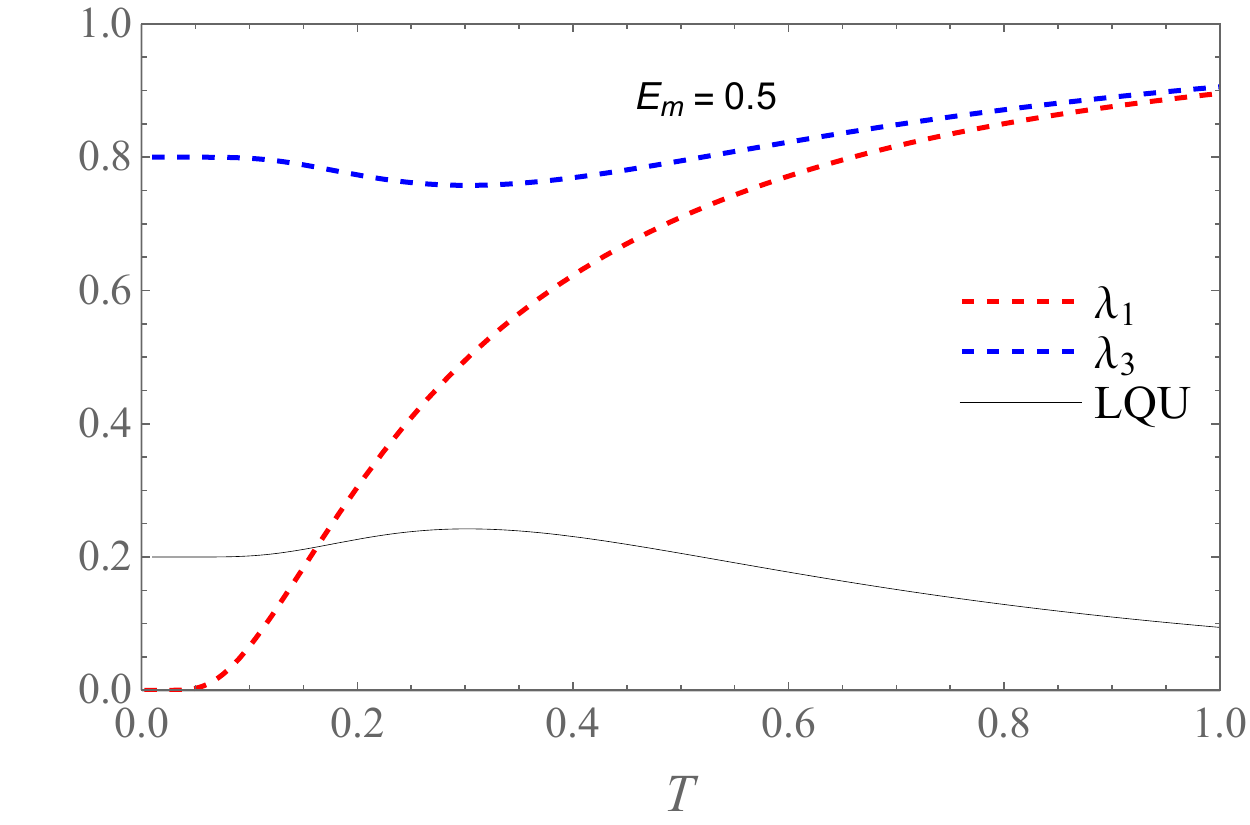}(a)\\
		\includegraphics[width=0.95\linewidth]{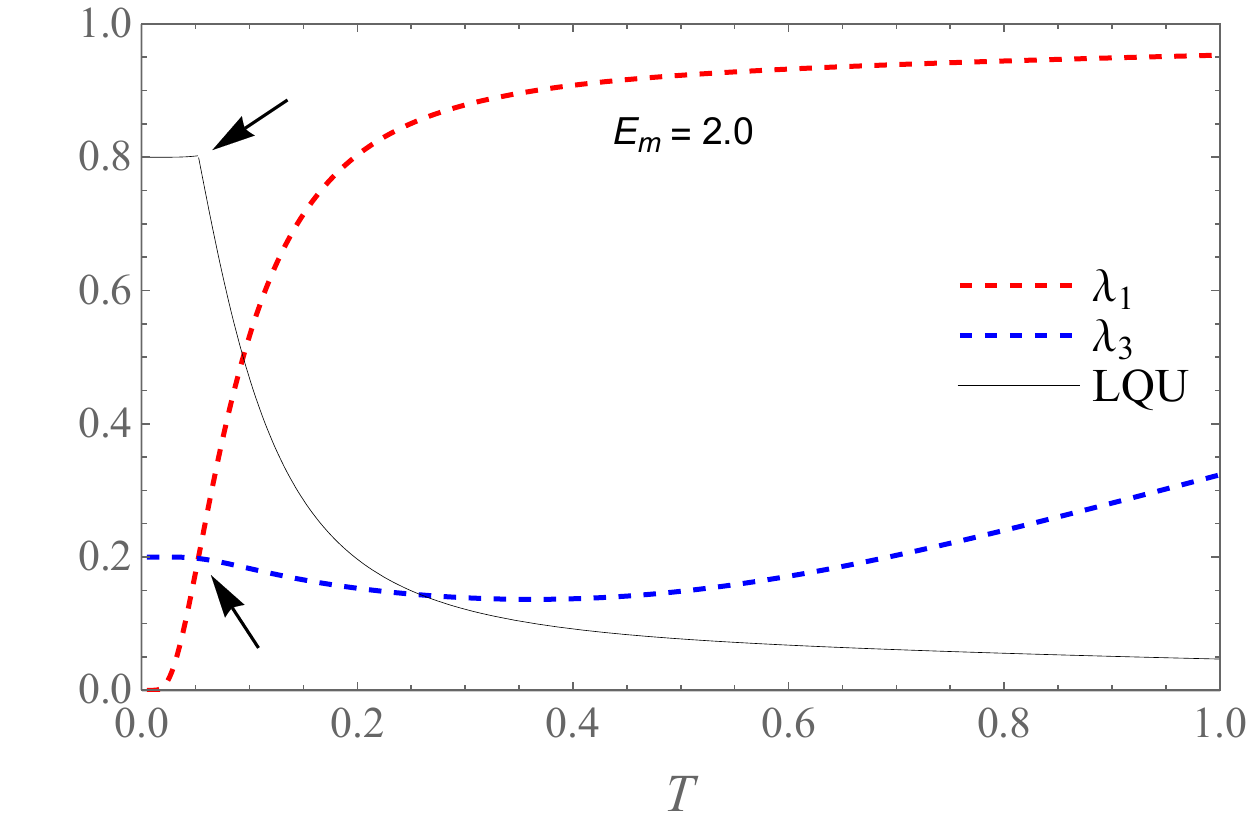}(b)
	\caption{LQU and temperature dependence of the eigenvalues $\lambda_{1}$ and $\lambda_{3}$ for (a) $E_{m}=0.5$ and (b) $E_{m}=2$ with $E_{j}=1$. }
	\label{ej02}
\end{figure}

From Eq. \eqref{lqu4}, it is clear that the temperature dependence of LQU is determined by $\lambda_{1}$ and $\lambda_{3}$. Particularly, the LQU and the mentioned eigenvalues at $E_{m}=1$ and different values of $E_{j}$ are shown in figure \ref{ej01}. At small $E_{j}=0.5$ [see  figure \ref{ej01}(a)], the first eigenvalue $\lambda_{1}$ goes towards one when $T\rightarrow 1$, hence, the LQU tends to zero. At a certain temperature,  we have a crossover between two eigenvalues $\lambda_{1}$ and $\lambda_{3}$ and the LQU demonstrates a sharp maximum at this temperature. At high Josephson energy, e.g., $E_{j}=2$ [see figure \ref{ej01}(b)], the two eigenvalues are separated and the third eigenvalue $\lambda_{3}$ is the maximum with respect to $\lambda_{1}$. So, the LQU is determined by temperature dependence of $\lambda_{3}$ only.

As seen in figure \ref{ej02}, the similar situation is occur if we change the coupling energies. At small $E_{m}=0.5$ [see figure \ref{ej02}(a)], the two eigenvalues are distinct from each other, and LQU is determined only by $\lambda_{3}$. However, when $E_{m}=2$ [see figure \ref{ej02}(b)], the LQU undergoes a sudden transition in the energy crossover between the two eigenvalues $\lambda_{1}$ and $\lambda_{3}$ at a specific temperature.

\subsection{Thermal LQU under decoherence}
Quantum correlations are known to be of utmost importance in the realm of quantum information processing and quantum communication protocols. However, the interaction of quantum systems with their surroundings results in a mechanism termed as decoherence, leading to the loss of quantum coherence and the disintegration of subtle quantum correlations. Consequently, the system tends to display more classical behavior and forfeits some of its distinct quantum characteristics. Decohering channels, which represent the physical processes responsible for decoherence in quantum systems, are frequently modeled as noise or perturbations that impact the quantum state, inducing it to become mixed or probabilistic rather than retaining a pure state.

In general, the dynamics of particles that interact autonomously with diverse surroundings is explicated by the solutions of the appropriate Born-Markov-Lindblad equations. For any bipartite quantum state that is initially set as $\varrho_{AB}$, the ultimate quantum state that arises under the influence of decohering channels, with the aid of the Kraus operator approach, can be determined by

\begin{equation}
\varepsilon\left(\varrho_{AB}\right):=\sum_{i, j} K_{i, j} \varrho_{AB} K_{i, j}^{\dagger},\label{Kraus}
\end{equation}
herein, it should be noted that the Kraus operators, denoted as $K_{i,j} = K_i \otimes K_j$, pertain to the one-qubit quantum channels represented by $K_i$ and $K_j$. It is imperative to emphasize that these operators must adhere to the closure condition whereby $\sum_{i, j} K_{i, j}^{\dagger} K_{i, j}=I$ at all times.

In this particular section, we shall examine the quantum correlation attributes that are associated with our thermal state \eqref{thermal}, which has been subject to decoherence effects. Hence, we posit that two superconducting qubits are autonomously coupled to three different decohering channels such as amplitude damping (AD), phase flip (PF), and phase damping (PD) \cite{n2000}.

\subsubsection{AD channel}
The AD decoherence channel is a well-known characterization of spontaneous emission that concerns the dissipation of energy from a quantum system. In the case of a single-qubit AD channel, its description can be succinctly formulated through the application of the following Kraus operators

\begin{equation}
K_1=\left(\begin{array}{cc}
1 & 0 \\
0 & \sqrt{1-p}
\end{array}\right), \quad K_2=\left(\begin{array}{cc}
0 & \sqrt{p} \\
0 & 0
\end{array}\right),\label{kad}
\end{equation}
where $p=1-\exp(-\kappa t)$ denotes the decoherence parameter with $p\in[0,1]$, and $\kappa$ is decay rate.
Now, through the substitution of Eqs. \eqref{thermal} and \eqref{kad} into Eq. \eqref{Kraus}, the new density matrix under AD effect can be represented as
\begin{equation}
\varrho_{T}^{AD}=\left(
\begin{array}{cccc}
 \delta_1 & 0 & 0 & c-c p \\
 0 & \delta_2 & d-d p & 0 \\
 0 & d-d p & \delta_2 & 0 \\
 c-c p & 0 & 0 & \delta_3 \\
\end{array}
\right),
\end{equation}
where $\delta_1=a^{+}+p (2 b+a^{-} p)$, $\delta_2=-(p-1) (b+a^{-} p)$, and $\delta_3=a^{-} (p-1)^2$.
Then, we can obtain an explicit expression for LQU \eqref{lqu4} as follows

\begin{equation}\label{lquAD}
\textmd{LQU}(\varrho_T^{AD})=1- \max\{\lambda_1^{AD},  \lambda_3^{AD}\}.
\end{equation}
where

\begin{align}
 \lambda_1^{AD}=&\left(\sqrt{\mathcal{A}_1}+\sqrt{\mathcal{A}_2}\right)\left(\sqrt{\mathcal{A}_3}+\sqrt{\mathcal{A}_4}\right)\nonumber\\
 &+\frac{4| (c-c p)(d-d p)|}{\left(\sqrt{\mathcal{A}_1}+\sqrt{\mathcal{A}_2}\right)\left(\sqrt{\mathcal{A}_3}+\sqrt{\mathcal{A}_4}\right)}, \nonumber\\
\lambda_3^{AD}=&\frac{1}{2}\bigg[\left(\sqrt{\mathcal{A}_1}+\sqrt{\mathcal{A}_2}\right)^2+\left(\sqrt{\mathcal{A}_3}+\sqrt{\mathcal{A}_4}\right)^2
\nonumber\\&+\frac{(\delta_3-\delta_1)^2 - 4|c-c p|^2}{\left(\sqrt{\mathcal{A}_1}+\sqrt{\mathcal{A}_2}\right)^2}-\frac{4|d-dp|^2}{\left(\sqrt{\mathcal{A}_3}+\sqrt{\mathcal{A}_4}\right)^2}\bigg],\nonumber
\end{align}
with $\mathcal{A}_{1,2}=\frac{1}{2}\left[\delta_1+\delta_3 \pm \sqrt{(\delta_1-\delta_3)^2 + 4|c-c p|^2}\right]$
and
$\mathcal{A}_{3,4}=\delta_2 \pm |d-d p|.$

\subsubsection{PF channel}
In this particular channel, a dynamic operates with a probability of $p$ upon the initial state, giving rise to a unitary transformation characterized by $\sigma_z$, and consequently, a phase flip occurs. A typical representation of PF decoherence channel for a single-qubit can be defined by means of the following Kraus operators
\begin{equation}
K_1=\left(\begin{array}{cc}
\sqrt{p} & 0 \\
0 & \sqrt{p}
\end{array}\right), \quad K_2=\left(\begin{array}{cc}
\sqrt{1-p} & 0 \\
0 & -\sqrt{1-p}
\end{array}\right).\label{kpf}
\end{equation}

According to the previous method, the new density matrix based on our thermal state \eqref{thermal} can be obtained as

\begin{equation}
\varrho_{T}^{PF}=\left(
\begin{array}{cccc}
 a^{+} & 0 & 0 & c (1-2 p)^2 \\
 0 & b & d (1-2 p)^2 & 0 \\
 0 & d (1-2 p)^2 & b & 0 \\
 c (1-2 p)^2 & 0 & 0 & a^{-} \\
\end{array}
\right),
\end{equation}
and for LQU, we have

\begin{equation}\label{lquPF}
\textmd{LQU}(\varrho_T^{PF})=1- \max\{\lambda_1^{PF},  \lambda_3^{PF}\}.
\end{equation}
where

\begin{align}
 \lambda_1^{PF}=&\left(\sqrt{\mathcal{B}_1}+\sqrt{\mathcal{B}_2}\right)\left(\sqrt{\mathcal{B}_3}+\sqrt{\mathcal{B}_4}\right)\nonumber\\
 &+\frac{4|c d(1-2 p)^4|}{\left(\sqrt{\mathcal{B}_1}+\sqrt{\mathcal{B}_2}\right)\left(\sqrt{\mathcal{B}_3}+\sqrt{\mathcal{B}_4}\right)},\nonumber\\
\lambda_3^{PF}=&\frac{1}{2}\bigg[\left(\sqrt{\mathcal{B}_1}+\sqrt{\mathcal{B}_2}\right)^2+\left(\sqrt{\mathcal{B}_3}+\sqrt{\mathcal{B}_4}\right)^2
\nonumber\\&+\frac{(a^{-}-a^{+})^2 - 4|(1-2 p)^2|^2}{\left(\sqrt{\mathcal{B}_1}+\sqrt{\mathcal{B}_2}\right)^2}-\frac{4|d(1-2 p)^2|^2}{\left(\sqrt{\mathcal{B}_3}+\sqrt{\mathcal{B}_4}\right)^2}\bigg],\nonumber
\end{align}
with $\mathcal{B}_{1,2}=(a^{+}+a^{-} \pm \sqrt{(a^{+}-a^{-})^2 + 4|c(1-2 p)^2|^2})/2$ and $\mathcal{B}_{3,4}=b \pm |d(1-2 p)^2|$.

\subsubsection{PD channel}
As is commonly understood, a PD decoherence channel pertains to a quantum noise that results in a loss of quantum phase information, yet does not result in a loss of energy. The Kraus operators for a single-qubit PD channel are explicitly provided by

\begin{equation}
K_1=\left(\begin{array}{cc}
1 & 0 \\
0 & \sqrt{1-p}
\end{array}\right), \quad K_2=\left(\begin{array}{cc}
0 & 0 \\
0 & \sqrt{p}
\end{array}\right).\label{kpd}
\end{equation}

After passing the thermal state \eqref{thermal} through the PD decoherence channel, the resultant density matrix is expressed as follows
\begin{equation}
\varrho_{T}^{PD}=\left(
\begin{array}{cccc}
 a^{+} & 0 & 0 & c-c p \\
 0 & b & d-d p & 0 \\
 0 & d-d p & b & 0 \\
 c-c p & 0 & 0 & a^{-} \\
\end{array}
\right).
\end{equation}

Next, one can obtain the analytical expression of LQU for $\varrho_{T}^{PD}$ as
\begin{equation}\label{lquPD}
\textmd{LQU}(\varrho_T^{PD})=1- \max\{\lambda_1^{PD},  \lambda_3^{PD}\}.
\end{equation}
where
\begin{align}
\lambda_1^{PD}=&\left(\sqrt{\mathcal{C}_1}+\sqrt{\mathcal{C}_2}\right)\left(\sqrt{\mathcal{C}_3}+\sqrt{\mathcal{C}_4}\right)\nonumber\\&+\frac{4| (c-c p)(d-d p)|}{\left(\sqrt{\mathcal{C}_1}+\sqrt{\mathcal{C}_2}\right)\left(\sqrt{\mathcal{C}_3}+\sqrt{\mathcal{C}_4}\right)}, \nonumber\\
\lambda_3^{PD}=&\frac{1}{2}\bigg[\left(\sqrt{\mathcal{C}_1}+\sqrt{\mathcal{C}_2}\right)^2+\left(\sqrt{\mathcal{C}_3}+\sqrt{\mathcal{C}_4}\right)^2
\nonumber\\&+\frac{(a^{-}-a^{+})^2 - 4|c-c p|^2}{\left(\sqrt{\mathcal{C}_1}+\sqrt{\mathcal{C}_2}\right)^2}-\frac{4|d-d p|^2}{\left(\sqrt{\mathcal{C}_3}+\sqrt{\mathcal{C}_4}\right)^2}\bigg],\nonumber
\end{align}
with $\mathcal{C}_{1,2}=(a^{+}+a^{-} \pm \sqrt{(a^{+}-a^{-})^2 + 4|c-cp|^2})/2$ and $\mathcal{C}_{3,4}=b \pm |d-dp|$.

\begin{figure}[t]
	\centering
		\includegraphics[width=0.88\linewidth]{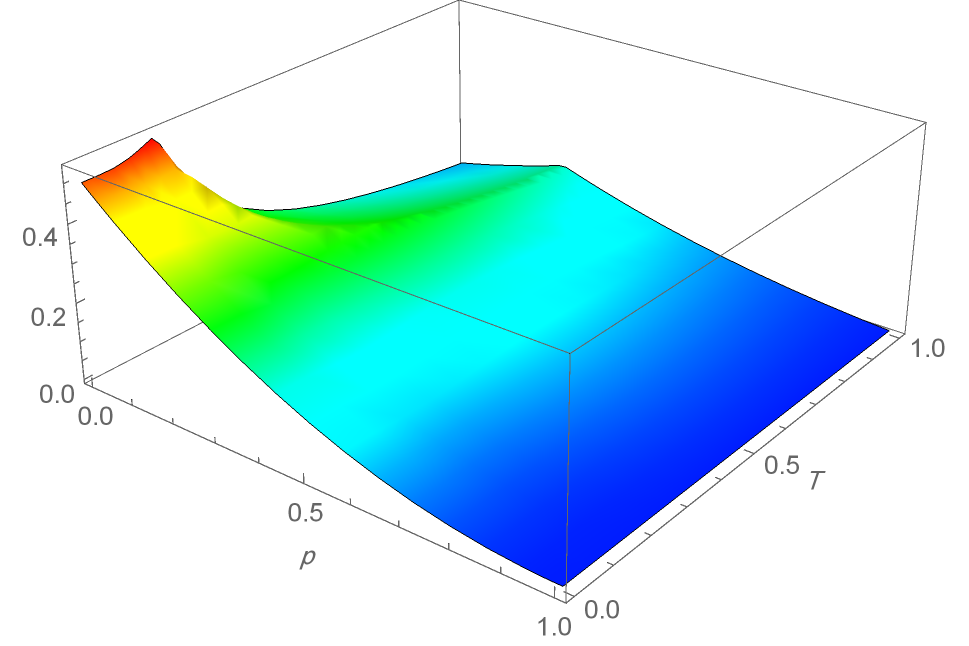}(a)\\
		\includegraphics[width=0.88\linewidth]{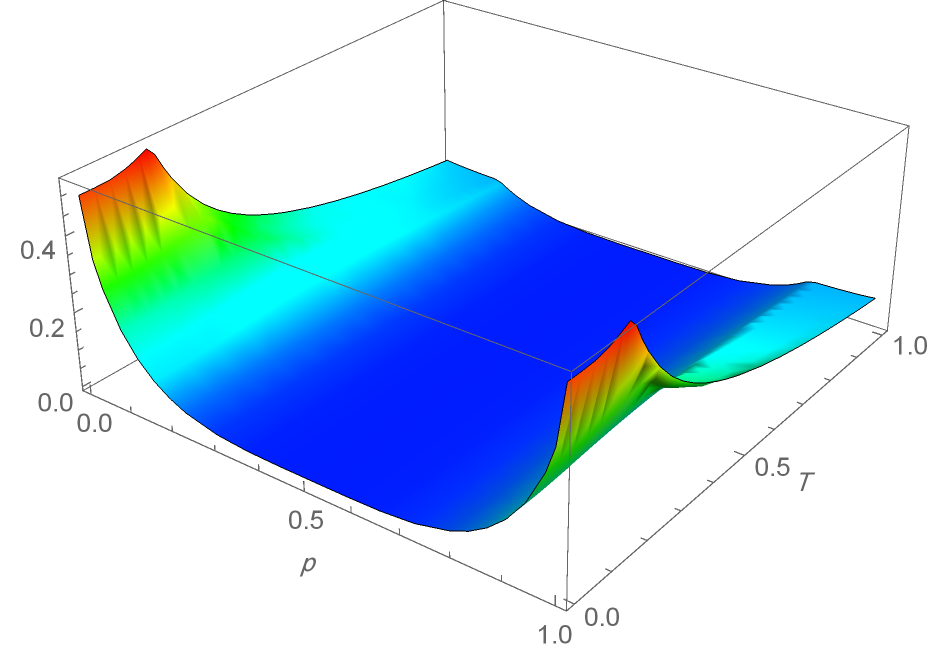}(b)\\
        \includegraphics[width=0.88\linewidth]{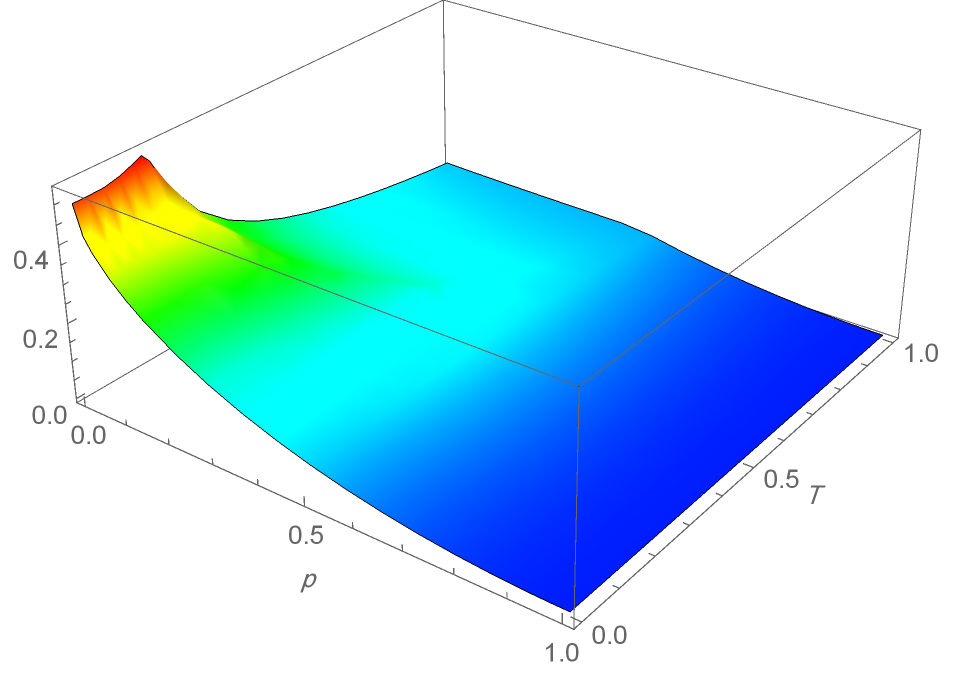}(c)
	\caption{LQU versus $p$ and $T$ for (a) AD channel, (b) PF channel, and (c) PD channel with $E_{m}=E_{j}=1$. }
	\label{fig6}
\end{figure}

\begin{figure}[htb]
	\centering
		\includegraphics[width=0.75\linewidth]{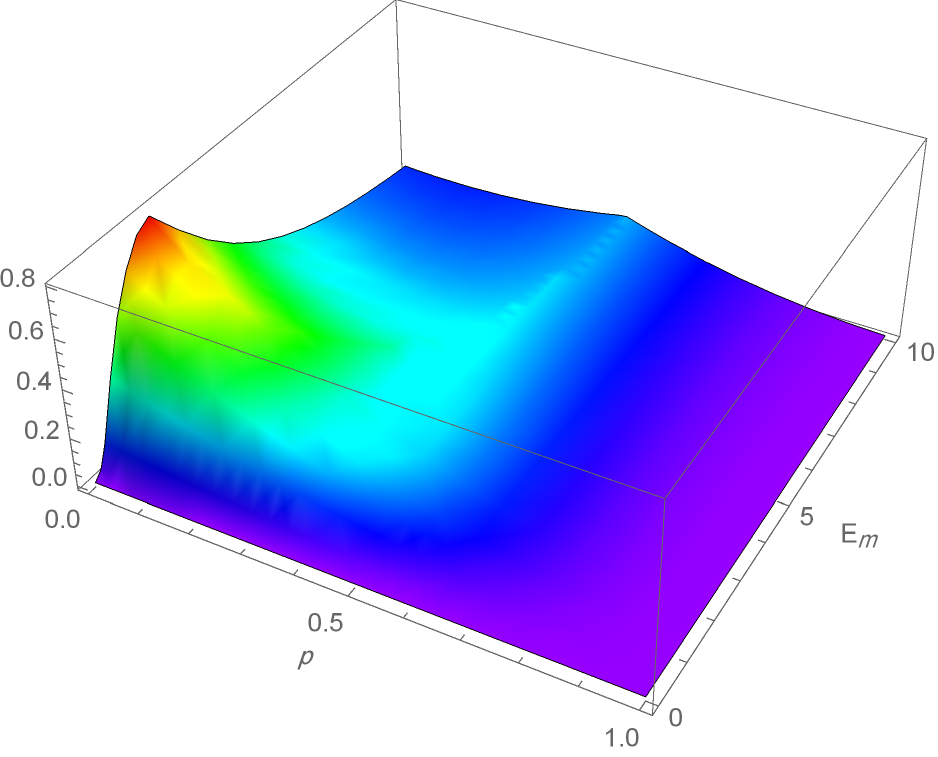}(a)\\
		\includegraphics[width=0.75\linewidth]{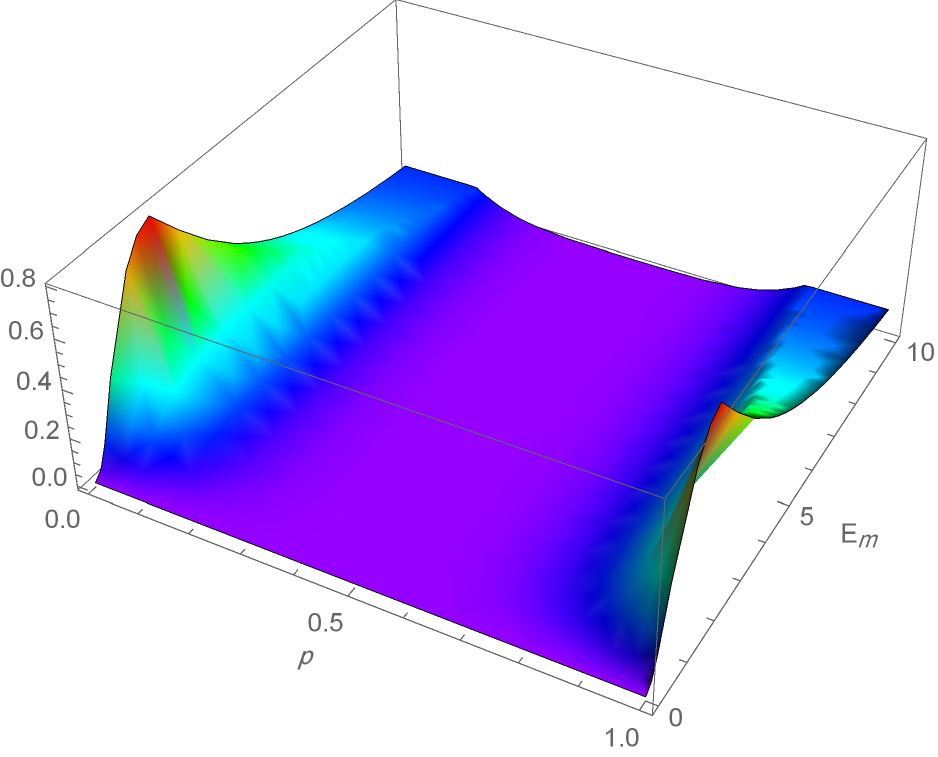}(b)\\
        \includegraphics[width=0.75\linewidth]{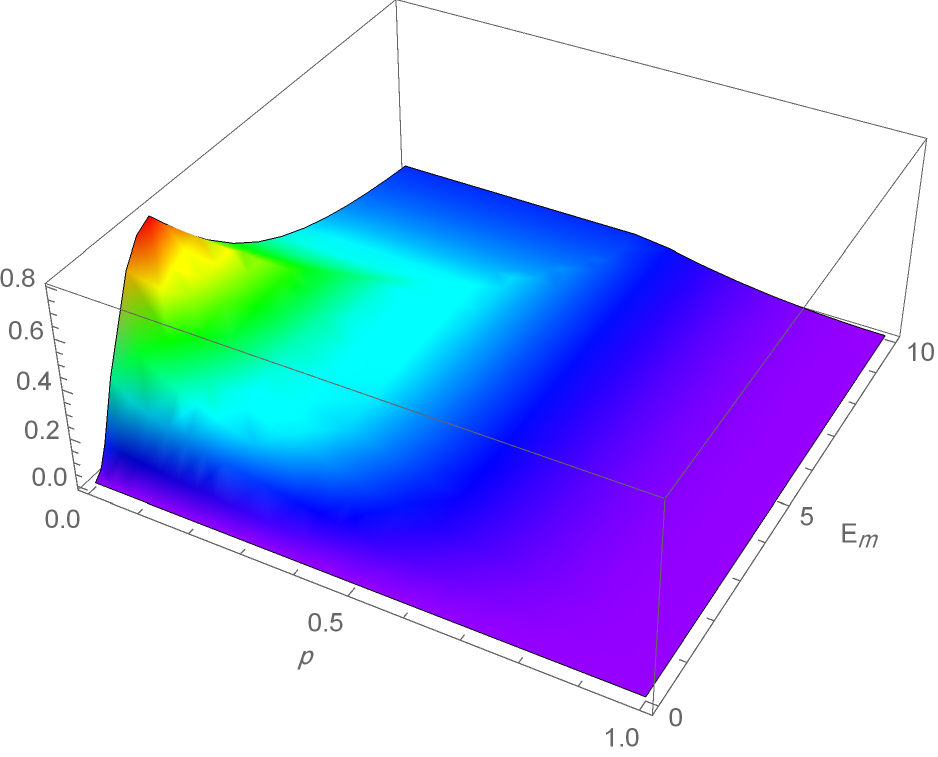}(c)
	\caption{LQU versus $p$ and $E_{m}$ for (a) AD channel, (b) PF channel, and (c) PD channel with $E_{j}=1$ and T=0.05. }
	\label{fig7}
\end{figure}

\begin{figure}[htb]
	\centering
		\includegraphics[width=0.77\linewidth]{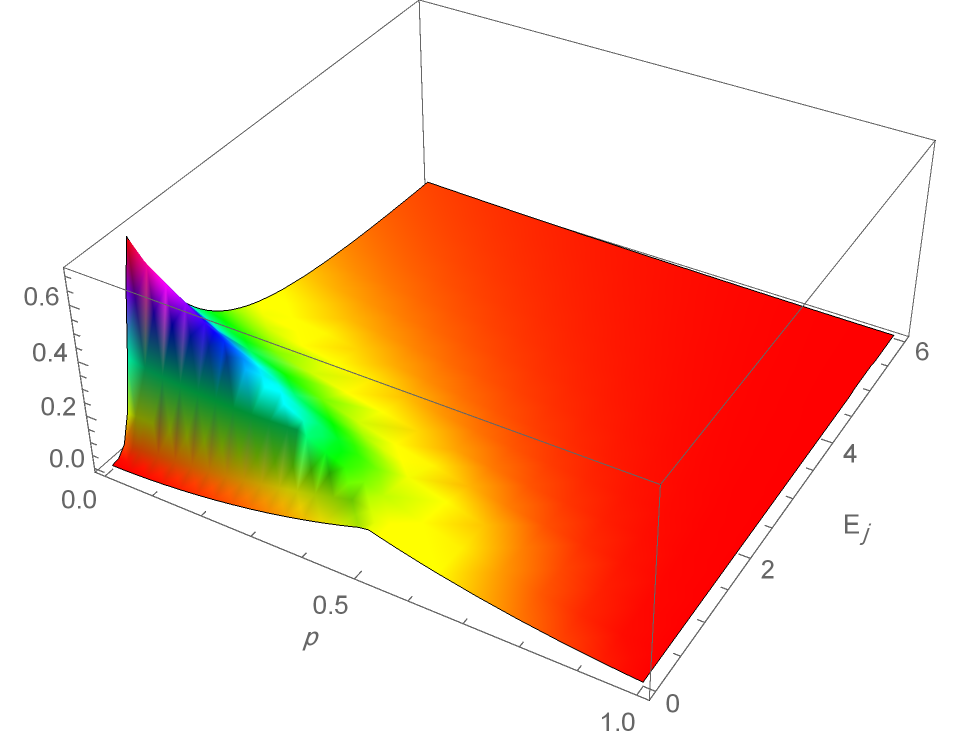}(a)\\
		\includegraphics[width=0.77\linewidth]{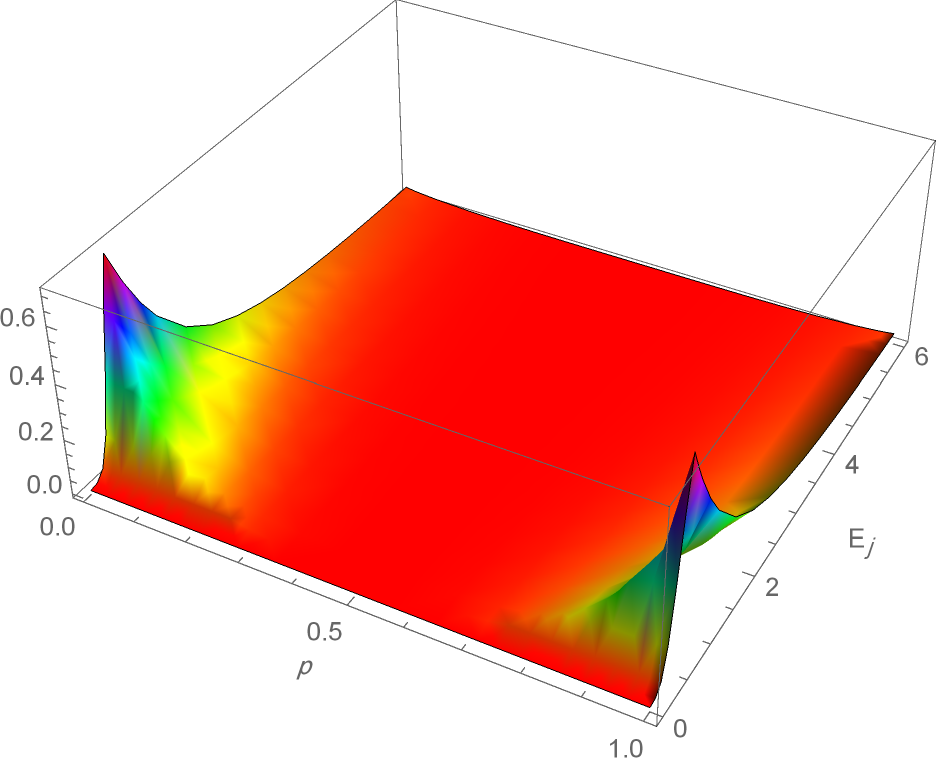}(b)\\
        \includegraphics[width=0.77\linewidth]{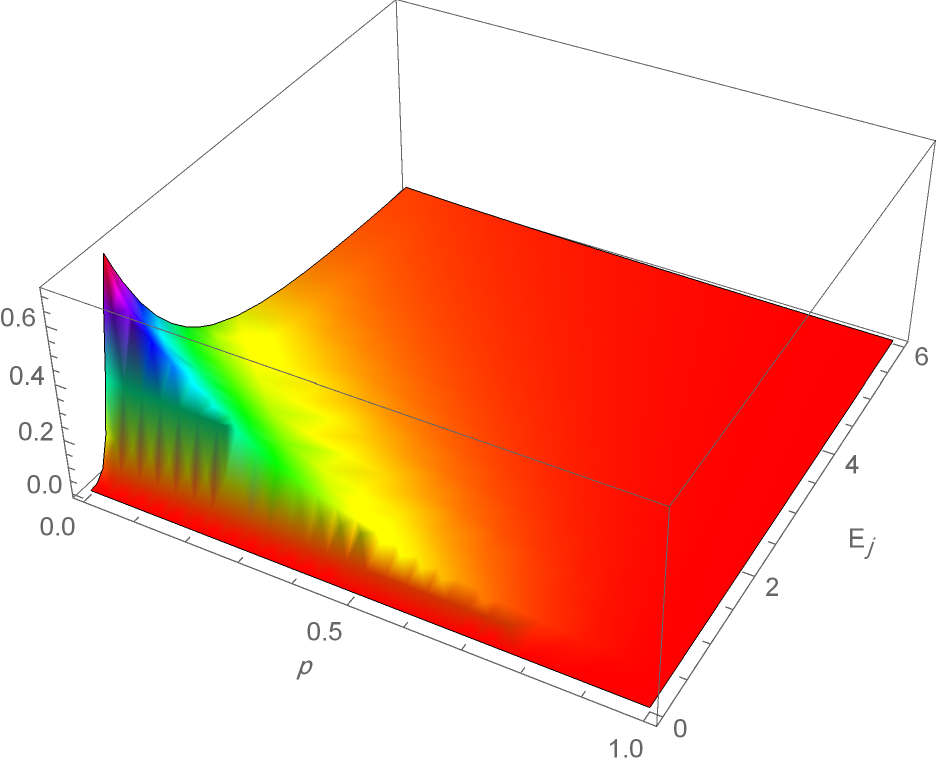}(c)
	\caption{LQU versus $p$ and $E_{j}$ for (a) AD channel, (b) PF channel, and (c) PD channel with $E_{m}=1$ and T=0.05.}
	\label{fig8}
\end{figure}

\begin{figure}[htb]
	\centering
		\includegraphics[width=0.90\linewidth]{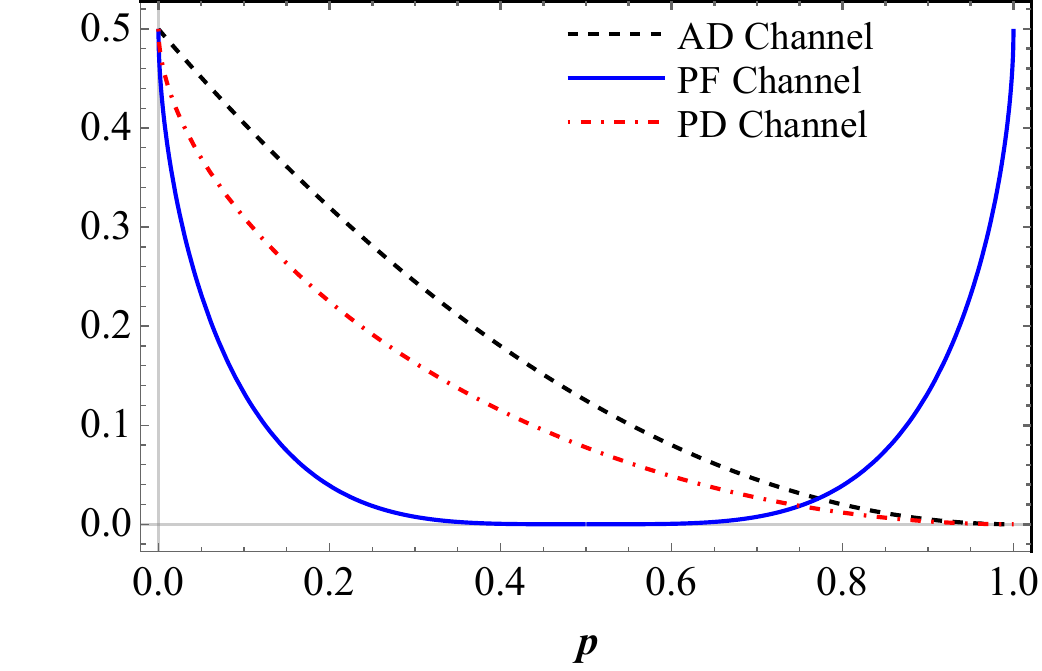}(a)\\
		\includegraphics[width=0.90\linewidth]{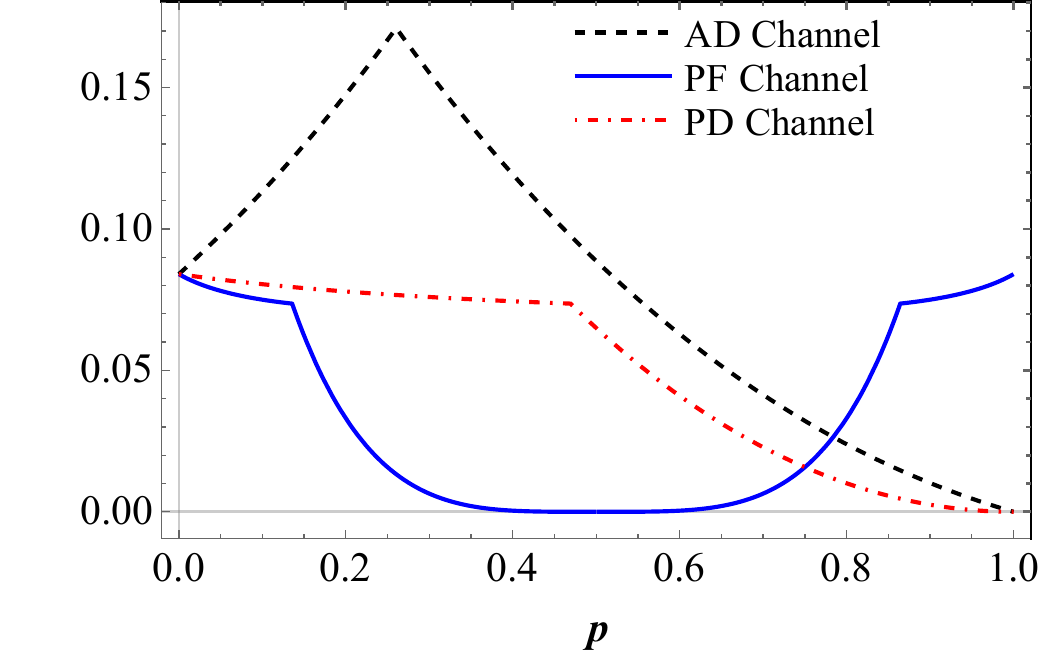}(b)
	\caption{LQU versus $p$ for (a) $T=0.05$ and (b) $T=1$ with $E_{m}=E_{j}=1$. }
	\label{fig9}
\end{figure}

Now, let us analyze the evolution of LQU for our thermal TQS system under three types of decohering channels as functions of temperature $T$, Josephson energy $E_j$,  coupling energy $E_m$, and decoherence parameter $p$ using the final formula for AD \eqref{lquAD}, PF \eqref{lquPF}, and PD \eqref{lquPD} channels.

The LQU has been plotted in three dimensions as a function of $p$ and the other parameters in Figs. \ref{fig6} to \ref{fig8} for different types of decohering channels.   In general, the evolution of LQU is the same under AD and PD channels such that LQU decays when $p$ increases and it has a symmetric form around $p=0.5$ in PF channel. However, by comparing Figs. \ref{fig6}-\ref{fig8} to Fig. \ref{fig9} in which we have plotted LQU versus $p$ for specific values $E_m$ and $E_j$, it is clear that under the PD channel, LQU goes to zero more rapidly than AD channel and hence the quantum correlation between two superconducting qubits is more robust under AD channel as shown in Fig. \ref{fig9}(a) at low temperatures. Besides, at high temperature ($T=1$), LQU begins from less value than at low temperature and under the AD channel, LQU rises to get a maximal value and then it reduces with growing $p$. But, under the PD channel, quantum correlation approximately remains unchanged for weak values of $p$ and then it decreases more rapidly than under the AD channel, as shown in Fig. \ref{fig9}(b).

In the case of the PF channel, LQU decreases till it vanishes completely around $p=0.5$ and it revives to get its previous maximal value as plotted in Figs. \ref{fig6}(b), \ref{fig7}(b), and \ref{fig8}(b). We emphasize that when $p=0$, all the results are the same as the outcomes in the previous section.

\section{Significance and outlook}\label{sec:5}
Superconducting qubits represent a preeminent platform for constructing quantum computers \cite{qc1,qc2,qc3} and quantum batteries \cite{Huck2022,Dou2023}. Quantum computing endeavors to leverage quantum correlations to resolve specific problems with an exponential speed-up relative to classical computers. Nevertheless, the occurrence of decoherence from interactions with the surrounding environment poses a significant obstacle to preserving and manipulating quantum information. Examining quantum correlations in a thermal TQS system subjected to decohering channels may furnish valuable insights into the effects of environmental interactions on quantum computation, and may facilitate the development of more robust and secure quantum communication protocols. Progress in comprehending and regulating quantum correlations in such systems may culminate in the creation of more robust and fault-tolerant quantum computers. Furthermore, our research may inform the design of superconducting qubit devices that are more resistant to environmental noise, a crucial prerequisite for the realization of practical quantum technologies.

The investigation of quantum correlations in a thermal TQS system subjected to decohering channels is a significant avenue of research with broad implications for the field of quantum information science and technology. The combination of theoretical analysis and experimental validation offers a promising avenue to further our understanding of quantum phenomena and enhance the performance of quantum devices. As the field of quantum technologies continues to advance, such research is likely to remain at the forefront of cutting-edge developments.

On one hand, superconducting qubits are among the leading candidates for building practical quantum computers due to their scalability, low error rates, and relatively simple fabrication process. However, there exist many challenges that must be overcome before large-scale quantum computers that rely on superconducting qubits can become a reality. On the other hand, superconducting qubits possess immense significance in the realm of quantum batteries owing to their quantum coherence properties, entanglement capabilities, scalability, and potential for quantum parallelism. As research in this field progresses, superconducting qubits are expected to play a pivotal role in the development of efficient and advanced quantum battery technologies.

\section{Conclusion}\label{sec:6}
We have studied the thermal LQU as a measure of quantum correlations in a TQS system.  Indeed, we have demonstrated the temperature dependence of LQU at different values of Josephson and coupling energies. A maximum on this dependence is found, which is explained through the temperature dependence of the largest eigenvalue of a symmetric matrix determining LQU.  We have shown that variation in the Hamiltonian parameters allows manipulation and control of the LQU. Specifically, the LQU of the considered system is greatly enhanced when the coupling energy is high while Josephson energy and temperature are low.

Moreover, we conducted both analytical and numerical analyzes to examine the LQU within the system of interest, subject to three distinct types of decohering channels. Under the framework of the PF channel, the quantum correlation properties exhibited notable distinctions when compared to those manifested under the AD and PD channels. Notably, the evolution of LQU exhibited identical characteristics in AD and PD channels, whereby an increase in $p$ resulted in the decay of LQU, however, it displayed a symmetrical form around $p=0.5$ under the PF channel.

Thereby, our results provide a way to enhance the quantum correlation degree of the system by manipulating the coupling and Josephson energies which can be realized experimentally, similar to what is followed in Ref. \cite{Shaw2009} for thermal entanglement of TQS system.

\vspace{20pt}

\section*{Acknowledgments}
The authors are grateful to I. R. Rahmonov and A. A. Mazanik for fruitful discussion about the results of this paper.  Numerical simulations were funded by Project No. 22-71-10022 of the Russian Scientific Fund. M. R. Pourkarimi thanks Prof. Yu. M. Shukrinov and the BLTP, JINR, Dubna in Russia for their generous hospitality where a part of this work was done. S. H. acknowledges the support of Semnan University.
\\
\section*{Disclosures}
The authors declare that they have no known competing financial interests.
\\
\section*{Data availability}
No datasets were generated or analyzed during the current study.
\\
\section*{ORCID iDs}
\noindent M. R. Pourkarimi \\ \href{https://orcid.org/0000-0002-8554-1396}{https://orcid.org/0000-0002-8554-1396}\\
S. Haddadi \\ \href{https://orcid.org/0000-0002-1596-0763}{https://orcid.org/0000-0002-1596-0763}\\
M. Nashaat \\ \href{https://orcid.org/0000-0003-2534-2544}{https://orcid.org/0000-0003-2534-2544}\\
K. V. Kulikov \\ \href{https://orcid.org/0000-0003-0744-8847}{https://orcid.org/0000-0003-0744-8847}\\
Yu. M. Shukrinov \\ \href{https://orcid.org/0000-0003-2496-0375}{https://orcid.org/0000-0003-2496-0375}\\



\end{document}